\documentclass[%
reprint,
superscriptaddress,
frontmatterverbose,
amsmath,amssymb,
pra,
]{revtex4-2}

\usepackage{amsfonts,amsmath,amssymb,amsthm}

\usepackage{graphicx}
\usepackage{dcolumn}
\usepackage{bm}

\usepackage{hyperref}

\hypersetup{
	pdfborder={0 0 0}
	pdfnewwindow=true,      
	colorlinks=true,       
	linkcolor=blue,          
	citecolor=blue,        
	filecolor=blue,      
	urlcolor=blue,          
}

\usepackage{color, soul}
\usepackage{amsmath}

\usepackage{lineno}
\makeatother

\newcommand{\be}{\begin{equation}}
\newcommand{\ee}{\end{equation}}

\usepackage{changes}

\usepackage{braket}

\newif\ifdraft
\drafttrue
\draftfalse

\begin{document}
  
\sethlcolor{yellow}
\title{Hybrid nano-domain structures of organic-inorganic perovskites from molecule-cage coupling effects}

\author{Ping Tuo}
\affiliation{AI for Science Institute, Beijing 100080, China}

\author{Lei Li}
\affiliation{Frontiers Science Center for Flexible Electronics, Xi'an Institute of Flexible Electronics (IFE) and Xi'an Institute of Biomedical Materials \& Engineering, Northwestern Polytechnical University, 127 West Youyi Road, Xi'an 710072, China}

\author{Xiaoxu Wang}
\affiliation{DP Technology, Beijing 100080, China}
\affiliation{AI for Science Institute, Beijing 100080, China}

\author{Jianhui Chen}
\affiliation{Fujian Science \& Technology Innovation Laboratory for Energy Devices of China, Ningde 352100, China}

\author{Zhicheng Zhong}
\email{zhong@nimte.ac.cn}
\affiliation{Key Laboratory of Magnetic Materials Devices \& Zhejiang Province Key Laboratory of Magnetic Materials and Application Technology, Ningbo Institute of Materials Technology and Engineering, Chinese Academy of Sciences, Ningbo 315201, China}
\affiliation{China Center of Materials Science and Optoelectronics Engineering, University of Chinese Academy of Sciences, Beijing 100049, China}

\author{Bo Xu}
\email{XuBo@catl-21c.com}
\affiliation{Fujian Science \& Technology Innovation Laboratory for Energy Devices of China, Ningde 352100, China}
\affiliation{Department of Physics, Laboratory of Computational Materials Physics, Jiangxi Normal University, Nanchang 330022, China}

\author{Fu-Zhi Dai}
\email{daifz@bjaisi.com}
\affiliation{AI for Science Institute, Beijing 100080, China}


\begin{abstract}
In hybrid perovskites, the organic molecules and inorganic frameworks exhibit distinct static and dynamic characteristics. Their coupling will lead to unprecedented phenomena, which have attracted wide research interests. In this paper, we employed Deep Potential molecular dynamics (DPMD), a large-scale MD simulation scheme with DFT accuracy, to study $\mathrm{FA/MAPbI_3}$ hybrid perovskites. A spontaneous hybrid nano-domain behavior, namely multiple molecular rotation nano-domains embedded into a single $\mathrm{[PbI_6]^{4-}}$ octahedra rotation domain, was firstly discovered at low temperatures. The behavior originates from the interplay between the long range order of molecular rotation and local lattice deformation, and clarifies the puzzling diffraction patterns of $\mathrm{FAPbI_3}$ at low temperatures. Our work provides new insights into the structural characteristics and stability of hybrid perovskite, as well as new ideas for the structural characterization of organic-inorganic coupled systems.

\end{abstract}

\maketitle

Organic inorganic Hybrid lead iodide perovskite has attracted great research interest due to its surging potential for photovoltaic (PV) applications \cite{chen2021advances}. Hybrid perovskites have the stoichemistry formula of $\mathrm{APbI_3}$, with A as organic molecule.
The structure of a hybrid perovskite consists of organic A-site cations dispersed in an inorganic periodic lattice.
The coupling effects between the organic molecules and the inorganic frameworks are concerned in many researches. For example, much attention has been paid to characterize the long range orientation order of A-site molecules \cite{PhysRevMaterials.2.073604,PhysRevLett.119.145501,PhysRevLett.122.225701,lahnsteiner2016room,beecher2016direct,bakulin2015real,leguy2015dynamics}, and it is well-known that the cubic perovskite phase consists of dynamically equilibrated tetragonal polycrystals \cite{zhao2020polymorphous,weber2018structural}, where the coupling effect prefers stable tetragonal structures locally instead of fully random fluctuations.
The coupling effect is not only physically intriguing, resulting in unprecedented phenomena, but also closely related to the stability issue of hybrid perovskites, which is the bottleneck limiting their photovoltaic applications \cite{meng2018addressing}.

The static and dynamic characteristics of molecular rotation and the $\mathrm{[PbI_6]^{4-}}$ octahedra rotation are distinct in time and length scales, yet highly correlated. This leads to great challenge in experiments as well as theoretical simulations to elucidate their behavior. For example, the $\beta-\gamma$ phase transition of $\mathrm{FAPbI_3}$ ($\mathrm{FA}$=formamidinium, $\mathrm{HC(NH_2)_2}$) was reported to be "re-entrant", where sharp divided diffraction peaks of the high temperature $\beta$ phase "fuse" into a single wider and weaker peak in the low temperature $\gamma$ phase \cite{fabini2016reentrant}. The "re-entrant" transition is a confusing phenomenon, which is unusual and questioned by other researchers \cite{weber2018phase}. 
Atomistic simulations are essential approaches to clarify the molecule-cage coupling effects of hybrid perovskites, and to evaluate their influences on the structural characteristics.
Due to the long reorientation time of A-site molecules, simulation methods with both high accuracy and high efficiency are desired, which has been made possible recently by machine learning assisted simulations \cite{zhang2021phase,Wu_PhysRevB_2021_v103_p24108,Tang_ActaMater_2021_v204_p116513,Dai_JMatSciTech_2020_v43_p168,Zhang_PhysPlas_2020_v27_p122704,Galib_Science_2021_v371_p921,Zeng_NatCommun_2020_v11_p5713,Yang_PhysRevLett_2021_v127_p80603,Wang_PhysRevX_2021_v11_p11006,Wang_ApplPhysLett_2019_v114_p244101,PhysRevLett.122.225701}. 
In this work, we adopted the Deep Potential molecular dynamics (DPMD) \cite{PhysRevLett.120.143001} to study polymorphic transitions of both $\mathrm{MAPbI_3}$ ($\mathrm{MA}$=methylamonium, $\mathrm{CH_3NH_3}$) and $\mathrm{FAPbI_3}$. 
The temperature dependent phase structures are clarified, where both materials go through phase transitions, from cubic ($\alpha$) to tetragonal ($\beta$), and to orthorhombic ($\gamma$), with temperature decreasing. In addition, nano-domain structures resulting from the molecule-cage coupling effect are firstly discovered in $\gamma$ phases, which will fuzz up the diffraction pattern and lead to difficulties in determining the exact crystal structure of $\gamma-\mathrm{FAPbI_3}$ by experimental methods.

\begin{figure}[htb]
    \centering
    \includegraphics[width=8.6 cm]{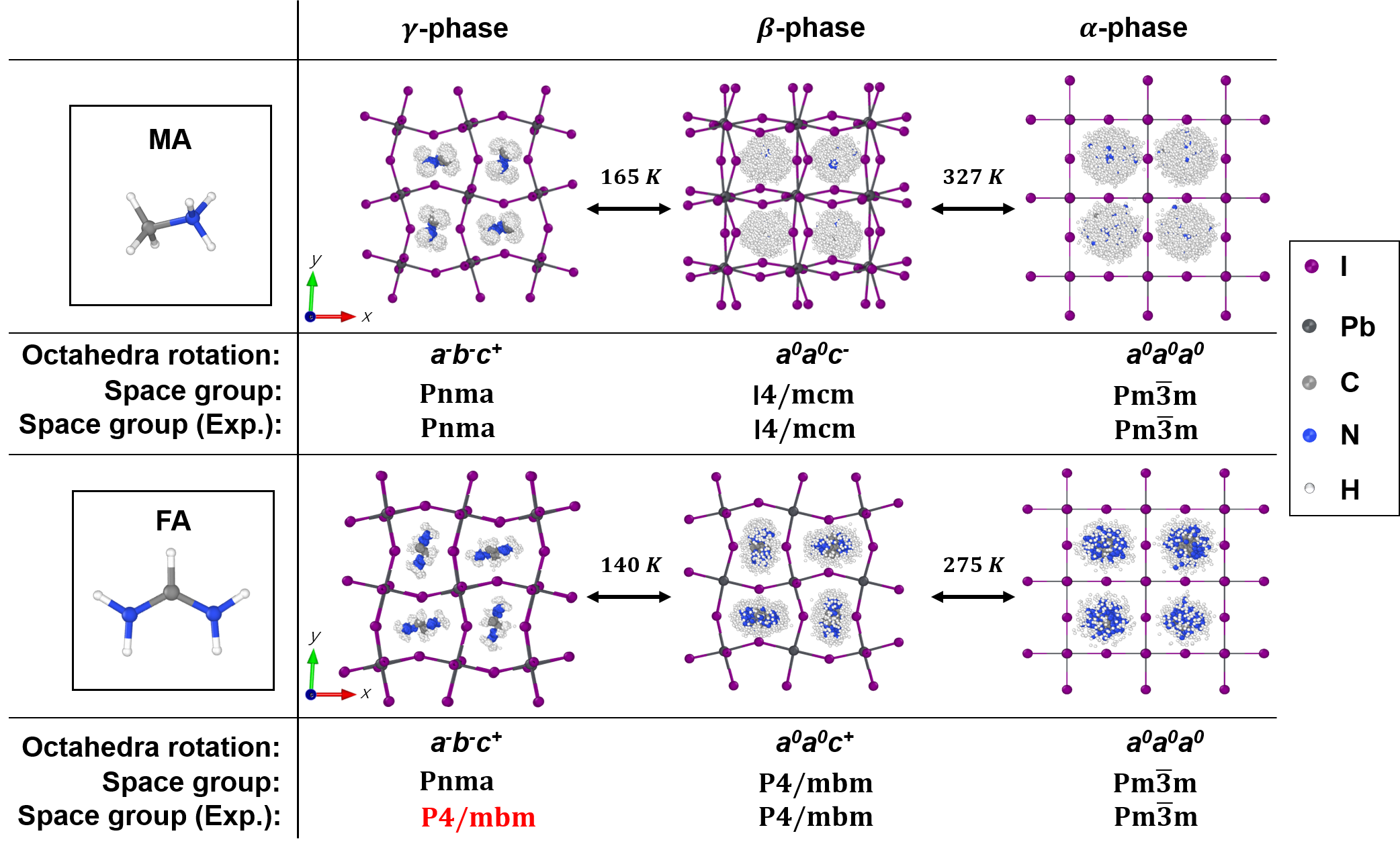}
    \caption{Overview of the temperature evolution of the perovskite lattice constructed according to symmetry, and time ensembles of organic A cation for $\mathrm{MAPbI_3}$ and $\mathrm{FAPbI_3}$ obtained from MD simulations. Experimental crystal structures and transition temperatures of $\mathrm{MAPbI_3}$ and $\mathrm{FAPbI_3}$ are taken from Weller et al. \cite{weller2015complete}, and Fabini et al. \cite{fabini2016reentrant} respectively.}
    \label{fig:1}
\end{figure}

\begin{figure*}[htb]
    \centering
    \includegraphics[width=17.2 cm]{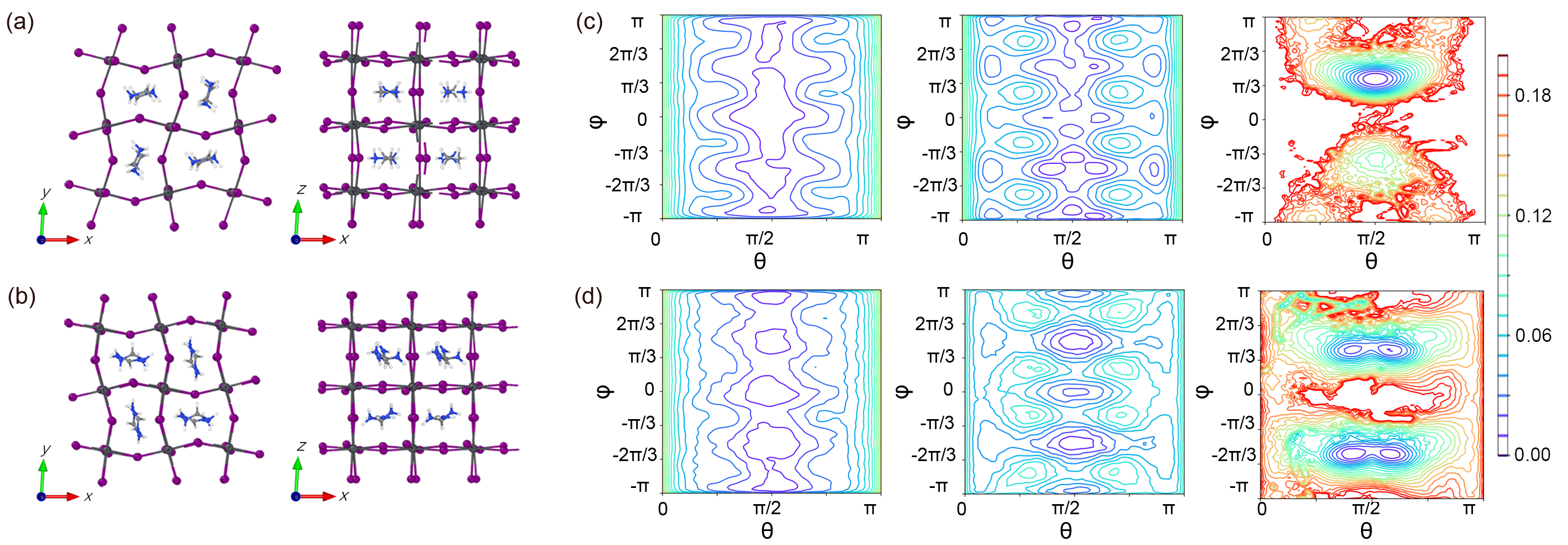}
    \caption{ $\gamma$ phases of (a) $\mathrm{MAPbI_3}$ and (b) $\mathrm{FAPbI_3}$ viewed from different directions. Free energy surface as a function of polar coordinates of (c) the C-N axis of MA and (d) the N-N axis of FA.}
    \label{fig:2}
\end{figure*}

\textsl{\textbf{Lattice symmetry.}} We begin our discussion by clarifying the lattice symmetry of each phase. The atomic configurations of the simulated $\alpha$, $\beta$ and $\gamma$ phase of $\mathrm{MAPbI_3}$ and $\mathrm{FAPbI_3}$ are illustrated in Figure \ref{fig:1}. The inorganic $\mathrm{[PbI_6]^{4-}}$ octahedra cages are constructed according to the simulated lattice symmetry, and indicated by their octahedra
rotation in Glazer notation \cite{glazer1972classification}, while the A-site molecules are illustrated in time ensembles obtained by MD simulations. 
In the high temperature $\alpha$ phase, A-site molecules rotate almost freely, and the dynamically distorted lattice adopts cubic symmetry (space group $\mathrm{Pm\bar{3}m}$) on average with equivalent tetragonal distortions \cite{Schwarzenbach2008Acta,beecher2016direct} and the octahedra rotation  $a^0a^0a^0$. 
In the medium temperature $\beta$ phase, $\mathrm{MAPbI_3}$ reduces to a body centered tetragonal lattice (space group $\mathrm{I4/mcm}$) with octahedra rotation $a^0a^0c^-$ \cite{ren2016orientation}, where the A-site $\mathrm{MA}$ molecules still rotate almost freely. In comparison, $\mathrm{FAPbI_3}$ reduces to a primitive tetragonal lattice (space group $\mathrm{P4/mbm}$) with octahedra rotation $a^0a^0c^+$ \cite{sun2017variable}, where the rotation of the A-site $\mathrm{FA}$ molecules are confined, especially for the N-N long axis \cite{fabini2017universal}. 
In the low temperature $\gamma$ phase, the A-site molecules are almost frozen. Both $\mathrm{MAPbI_3}$ and $\mathrm{FAPbI_3}$ reduces to orthorhombic lattice (space group $\mathrm{Pnma}$) with the $C_4$ symmetry lost and the octahedra rotation $a^-b^-c^+$  \cite{swainson2003phase,weller2015complete,lahnsteiner2016room,fabini2016reentrant,weber2016phase}. 
The frozen A-site molecules form a long range ordered pattern in the $\gamma$ phases. The relaxed configurations of $\gamma$ phases are shown in Figure \ref{fig:2}(a) and (b). In $\gamma-\mathrm{MAPbI_3}$, neighboring MA form an orthogonal pattern in the $xy$ plane, with the diagonally neighboring MA ordered ferroelectrically to each other. Thus, each plane should have a net electric polarization, which is canceled out by the antiferroelectric ordering of adjacent planes in the $z$ direction.  In comparison, the FA molecules in $\gamma-\mathrm{FAPbI}_3$ are also ordered orthogonally in the $xy$ plane, while the polar axis of neighboring FA molecules are orthogonal to each other, with a net dipole around zero.

\begin{figure}[htb]
    \centering
    \includegraphics[width=8.6 cm]{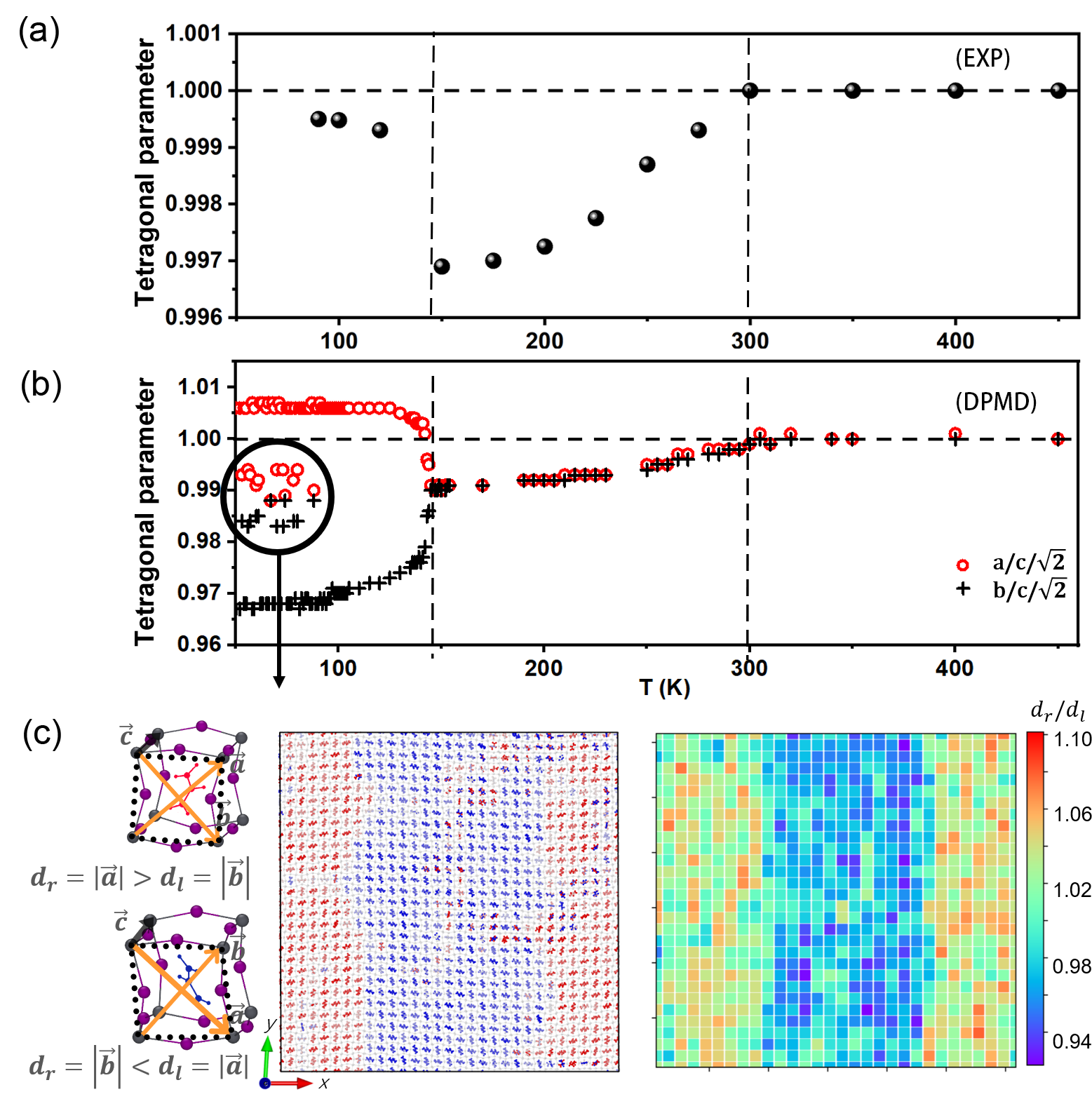}
    \caption{Phase transitions of $\mathrm{FAPbI_3}$. (a) Tetragonal lattice parameter ratio from synchrotron XRD experiments \cite{fabini2016reentrant}. 
    (b) Tetragonal lattice parameter ratio simulated by DPMD. The $\gamma$ phase is orthorhombic with lattice parameters $a\neq b$. For the circled points, $a\approx b$ due to multi-domains in the supercell. (c) Schematic display of the supercell with multi-domains. Multi-domains are colored by the orientation angle of $\mathrm{FA}$ molecules and the deformation of cages from a rectangular, respectively. The deformation of cages are indicated by the ratio $d_r/d_l$ in the figure. If $d_r/d_l=1$, the unit-cell is rectangular in the $xy$ plane; otherwise, the unit-cell is deformed.}
    \label{fig:3}
\end{figure}

\textsl{\textbf{Molecular rotation order.}} In experiments, the activation/deactivation of molecule rotation can be used as an indicator for phase transition \cite{fabini2017universal}. In simulations, the free energy surface (FES) of molecule reorientation can be used to visualize local symmetry, and hence to indicate structural transitions.
The on-the-fly probability enhanced sampling (OPES) simulation\cite{PhysRevX.10.041034} was adopted to ensure adequate sampling of molecular orientation, especially at lower temperatures. An MA (FA) molecule was randomly selected from a $4\times 4\times 4$ supercell, and the polar coordinates of its C-N (N-N) axis were used as collective variables. 

The FES of MA and FA rotation in different phases are given in Figure \ref{fig:2}(c) and (d), respectively. $\theta$ denotes the angle between the C-N axis of MA or the N-N axis of FA and $z$ direction, and $\phi$ is the angle in the $xy$ plane.
In $\alpha$ phase, the FESs show continuous low energy region in the $xy$ plane.
In $\beta$ phase, the FESs exhibit $C_4$ symmetry in the $xy$ plane. The valleys on the FESs correspond to the orientations lowest in energy, which reveal that face diagonal orientations are preferred by both MA and FA. Notably, in the case of MA, the $z$ direction of the tetragonal cell changes during MD simulations, as shown by the almost degenerate valleys at $\theta \approx 90^{\circ}$ and $\theta \approx 0^\circ$. When temperature is further lowered through $\beta-\gamma$ transition, $C_4$ symmetry is lost for both $\mathrm{MAPbI_3}$ and $\mathrm{FAPbI_3}$.  The sampled orientation of $\gamma-\mathrm{MAPbI_3}$ is insufficient in comparison with that of $\gamma-\mathrm{FAPbI_3}$, indicating that the molecule reorientation barrier of $\gamma-\mathrm{MAPbI_3}$ is higher than that of $\gamma-\mathrm{FAPbI_3}$, as confirmed and shown in Figure S7 and Figure S8. In $\gamma-\mathrm{FAPbI_3}$, the $\theta$ of the valleys deviate from $90^{\circ}$ by $\sim 16.1^{\circ}$, in agreement with experimental value of around $20^{\circ}$ reported by Weber et al. \cite{weber2018phase}.

\textsl{\textbf{Hybrid nano-domain structures.}} 
We start with the simulated variation of tetragonal parameters $a/c/\sqrt{2}$ and $b/c/\sqrt{2}$ of $\mathrm{FAPbI_3}$ with respect to temperature, as shown in Figure \ref{fig:3}.  As temperature is lowered through the $\alpha-\beta$ transition, the tetragonal parameters change from a value around $1$ to a value smaller than $1$, indicating distortion of the cell from a cubic one to a tetragonal one. When the temperature further decreases to induce the $\beta-\gamma$ transition, the parameters $a/c/\sqrt{2}$ and $b/c/\sqrt{2}$ are no longer equal to each other, indicating a tetragonal to orthorhombic transition. The predicted $\alpha-\beta$ and $\beta-\gamma$ transition points are around $300\mathrm{\ K}$ and $150 \mathrm{\ K}$ respectively, within $10 \mathrm{\ K}$ from the experimental values. Simulated variation of lattice parameters of $\mathrm{MAPbI_3}$ are provided in Figure S4, which also agree well with experiments.

In the work of Fabini et al. \cite{fabini2016reentrant}, a tetragonal cell model was proposed to fit the diffraction patterns of $\gamma-\mathrm{FAPbI_3}$, and the transition from $\beta$ to $\gamma$ phase with decreased temperature was named a ”re-entrant” transition. Later, Weber et al. \cite{weber2018phase} pointed out that the tetragonal model left some peaks unindexed. In our simulations, we also obtained some tetragonal-like $\gamma$ phase, where $a/c/\sqrt{2}$ and $b/c/\sqrt{2}$ are approximately equal to each other, as illustrated by the circled points in Figure \ref{fig:3}(b). These tetragonal-like points turn out to be the result of a non-uniform structure with nano-domains, as shown in Figure \ref{fig:3}(c). The domain structure is a hybrid one with the molecules rotated to multiple orientations while the octahedra rotation of inorganic framework remains $a^-b^-c^+$ . 
To characterize the domains, we define some parameters here. As explained before, the N-N axis of adjacent $\mathrm{FA}$ molecules are approximately parallel to the $x$ axis or the $y$ axis, and orthogonal to each other in the $xy$ plane. We denote the small angle between the N-N axis of horizontal molecules and $x$ axis, or the angle between the N-N axis of vertical molecules and $y$ axis as $\psi$, and use it to measure the molecular tilting from principle axes. $\psi$ takes a small value within $[-14^{\circ},14^{\circ}]$, where negative (positive) $\psi$ indicates counter-clock wise (clock wise) small angle rotation relative to the $x$ axis, or clock wise (counter-clock wise) small angle rotation relative to the $y$ axis. In Figure \ref{fig:3}(c), a molecule is colored in red if $\psi$ is negative, or colored in blue if $\psi$ is positive. Two distinct domains can be observed in Figure \ref{fig:3}(c) according to the small angle rotation relative to $x$ or $y$ axis. Variation of the molecular rotation in space leads to non-uniform local deformation of the lattice. To characterize this effect, we calculated the distribution of the ratio of the length of two diagonals $d_r/d_l$ in space, as illustrated in Figure \ref{fig:3}(c). Cells with $d_r/d_l > 1$ are colored in red, while cells with $d_r/d_l < 1$ are colored in blue. As shown in Figure \ref{fig:3}(c), domains distinguished by small angle molecular rotation $\psi$ and lattice deformation $d_r/d_l$ are coincident with each other, indicating the coupling nature between molecular rotation and deformation of their surrounding cells. The deformation of different nano-domains compensate each other, makes the supercell in Figure \ref{fig:3}(c) rectangular-like in the $xy$ plane, and thus induces nearly equal tetragonal parameters as the circled points in Figure \ref{fig:3}(b). The size of the domains are around $\sim 5\ \mathrm{nm}$. Similar analyses are also carried out for $\gamma-\mathrm{MAPbI_3}$ and nano-domains are also found, as given in Figure S4. The approximate characteristic domain size of $\gamma-\mathrm{MAPbI_3}$ is $\sim 2.5\ \mathrm{nm}$, around half the domain size of $\gamma-\mathrm{FAPbI_3}$.

To understand the mechanism driving the formation of the multi-domains, a modified Ising model with an auxiliary lattice is constructed, whose Hamiltonian is: 
\begin{equation}
    \label{eq:1}
    H = -J\sum_{<ij>} T_iT_j - J^\prime \sum_i T_iT^{\prime}_i,
\end{equation}
where pseudo-spins of the main lattice $T_i=+1$ or $-1$ corresponds to $d_r/d_l>1$ or $<1$ respectively, denoting the local deformation of the cell; pseudo-spins of the auxiliary lattice $T_i^{\prime}=+1$ or $-1$ corresponds to $\psi > 0$ or $\psi < 0$, respectively, denoting left or right small angle rotation of the molecule. Since the distances between neighboring molecules are around $6\mathrm{\ \mathring{A}}$, near the cutoff of the DP model, direct interactions between molecules are negligible. Instead, the molecules interact through the $\mathrm{[PbI_6]^{4-}}$ cages around them. Therefore, the domains can be viewed as a result of two kinds of couplings, the coupling between neighboring cages with coupling constant $J$, and the coupling between a cage and the molecule within it with coupling constant $J^{\prime}$. Standard Monte Carlo simulations were performed by using this model, starting from a randomly initialized lattice to simulate the effect of quenching in experiments. After one billion steps of simulation, the Ising lattice stabilizes at a mosaic pattern with separated "spin-up" and "spin-down" domains (see Figure S6). Characteristic size of the domains decrease with the ratio of the coupling constants $J^{\prime}/J$. The reorientation barrier for the molecule is an indicator of the coupling strength. The stronger the coupling is, the higher the barrier is.
The reorientation barrier for FA and MA are given in Figure S7, and compared with experimental results. The reorientation barrier for MA in the $\gamma$ phase is much higher than that for the long N-N axis of FA. Thus, the corresponding coupling constant ratio $J^{\prime}/J$ of $\gamma-\mathrm{MAPbI_3}$ is larger than that of $\gamma-\mathrm{FAPbI_3}$, and the domain size of $\gamma-\mathrm{MAPbI_3}$ should be smaller than that of $\gamma-\mathrm{FAPbI_3}$, agreeing with DPMD simulated results.
We speculate that the mechanism behind the larger $J^{\prime}/J$ of $\gamma-\mathrm{MAPbI_3}$ might be related to the higher hydrogen bonding energy of MA molecule with its surrounding cage. We calculated the hydrogen bonding energy using the method reported by Katrine et al. \cite{svane2017strong}. The hydrogen bonding energy per molecule of $\gamma-\mathrm{MAPbI_3}$ is around $-0.464 \mathrm{\ eV}$, around three times that of $\gamma-\mathrm{FAPbI_3}$ ($-0.144 \mathrm{\ eV}$).

To the best of our knowledge, the hybrid nano-domain structure in this work is a new structural phenomenon that has not been reported before. 
Multiple molecular orientation domains are embedded in a single orthorhombic inorganic domain with uniform $\mathrm{[PbI_6]^{4-}}$ octahedra rotation $a^-b^-c^+$. For $\mathrm{MAPbI_3}$, the inorganic framework transforms from a body-centered structure ($\beta$ phase with space group $\mathrm{I4/mcm}$ and octahedra rotation $a^0a^0c^-$) to a primitive structure ($\gamma$ phase with space group $\mathrm{Pnma}$ and octahedra rotation $a^-b^-c^+$), as shown in Figure \ref{fig:1}. The significant change in the framework will lead to prominent extra diffraction peaks. Therefore, identification of the $\beta$ and $\gamma$ phases of $\mathrm{MAPbI_3}$ is not difficult.
In contrast, in the case of $\mathrm{FAPbI_3}$, the structure of the $\mathrm{[PbI_6]^{4-}}$ framework of $\beta$ and $\gamma$ phases are both primitive structures. As a result, the diffraction pattern will not change substantially. In addition, the nano-domain structure leads to "fused" diffraction peaks due to the uneven interplaner spacing near domain boundaries.
"Fused" diffraction patterns were observed in experiments to validate our predictions, as shown in Figure 1 of reference \cite{fabini2016reentrant}, where separated sharp peaks of $\beta$ phase are "fused" into wide weak peaks in both $\alpha$ and $\gamma$ phases of $\mathrm{FAPbI_3}$.
Even though the temperature is the lowest, diffraction peaks of the $\gamma$ phase are the most wide and weak, indicating much higher internal stresses in the $\gamma$ phase.

To sum up, the lattice symmetry and molecule orientation of $\mathrm{MAPbI_3}$ and $\mathrm{FAPbI_3}$ are studied systematically over a wide temperature range of $\sim 50 \mathrm{\ K}-450 \mathrm{ \ K}$, emphasizing on the low temperature $\gamma$ phase. Both $\mathrm{MAPbI_3}$ and $\mathrm{FAPbI_3}$ go though polymorphic transitions from cubic ($\alpha$) to tetragonal ($\beta$) and to orthorhombic ($\gamma$) phase with decreasing temperature. In the $\gamma$ phase, neighboring A-site molecules are orthogonal to each other in the $xy$ plane and ordered antiferroelectrically in the $z$ direction. Hybrid nano-domain structures are firstly reported in the $\gamma$ phases of both perovskites, where nano-domains with multiple molecular rotations are embedded in a single $\mathrm{[PbI_6]^{4-}}$ octahedra rotation domain.
The underlying mechanism of the nano-domain structure is the molecule-cage coupling effect where the domain size depends on the coupling strength. This finding provides new insights into the structural characteristics and stability of hybrid perovskites, as well as new ideas for the structural characterization in organic-inorganic coupled systems.

\section*{Acknowledgments}
We are grateful to Xifan Wu of Temple University for valuable discussion.
This paper was supported by the National Key R\&D Program of China (Grants No. 2021YFA0718900), the Key Research Program of Frontier Sciences of CAS (Grant No. ZDBS-LY-SLH008), and the Science Center of the National Science Foundation of China (52088101).  

\bibliographystyle{elsarticle-num}  
\bibliography{references}

\end{document}